\documentclass[10pt,aps,pra,superscriptaddress,nofootinbib,twocolumn]{revtex4-2}

\usepackage{amsmath}
\usepackage[english]{babel}
\usepackage[T1]{fontenc}
\usepackage[colorlinks,citecolor=blue,linkcolor=blue]{hyperref}
\usepackage[utf8]{inputenc}
\usepackage{bbm}

\newcommand{\nocontentsline}[3]{}
\newcommand{\tocless}[2]{\bgroup\let\addcontentsline=\nocontentsline#1{#2}\egroup}

\usepackage[dvipsnames]{xcolor}
\usepackage{comment}

\usepackage{amsfonts}
\usepackage{amssymb}
\usepackage{amsthm}
\usepackage{braket}
\usepackage{dsfont}
\usepackage[dvipsnames]{xcolor}
\usepackage{graphicx}
\definecolor{nice}{RGB}{230,0,230}
\definecolor{darkgreen}{RGB}{50,190,50}
\usepackage{epigraph}

\theoremstyle{plain}
\newtheorem{theorem}{Theorem}

\newtheorem{axiom}{Principle}

\newtheorem{thmcorollary}{Corollary}[theorem]

\newtheorem{consequence}{Consequence}

\newtheorem*{example*}{Example}
\newtheorem{property}{Property}

\theoremstyle{definition}
\newtheorem{definition}{Definition}

\theoremstyle{remark}

\newcommand{\mc}[1]{\mathcal{#1}}
\DeclareMathOperator{\Tr}{Tr}

\usepackage{pifont}
\newcommand{\cmark}{\ding{51}}%
\newcommand{\xmark}{\ding{55}}%

\newcommand{\id}{1\!\!1}

\newcommand{\ralph}[1]{{\color{red} [#1]}}
\newcommand{\nuriya}[1]{{\color{Plum} [#1]}}

\DeclareRobustCommand{\rchi}{{\mathpalette\irchi\relax}}
\newcommand{\irchi}[2]{\raisebox{\depth}{$#1\chi$}} 

\begin{document}

\title{Ticking clocks in quantum theory}
\author{Ralph Silva}
\affiliation{Institute for Theoretical Physics, ETH Z\"urich, Wolfgang-Pauli-Strasse 27, 8093 Z\"urich, Switzerland}

\author{Nuriya Nurgalieva}
\affiliation{Institute for Theoretical Physics, ETH Z\"urich, Wolfgang-Pauli-Strasse 27, 8093 Z\"urich, Switzerland}

\author{Henrik Wilming}
\affiliation{Leibniz Universit\"at Hannover, Appelstraße 2, 30167 Hannover, Germany}
\date{\today}

\begin{abstract}
We present a derivation of the structure and dynamics of a ticking clock by showing that for finite systems a single natural principle serves to distinguish what we understand as ticking clocks from time-keeping systems in general. As a result we recover the bipartite structure of such a clock: that the information about ticks is a classical degree of freedom. We describe the most general form of the dynamics of such a clock, and discuss the additional simplifications to go from a general ticking clock to models encountered in literature. The resultant framework encompasses various recent research results despite their apparent differences. Finally, we introduce the information theory of ticking clocks, distinguishing their abstract information content and the actually accessible information.
\end{abstract}

\maketitle
\onecolumngrid
\begin{flushleft}
\setlength{\epigraphwidth}{4in}
\epigraph{Is time the wheel that turns, or the track it leaves behind?}{Robin Hobb, \textit{Fool's Errand}}
\end{flushleft}
\tocless{\section*{Content}}
\twocolumngrid
\renewcommand{\tocname}{\vspace{-26pt}}
\tableofcontents


\newpage

\twocolumngrid

\section{Introduction}
\begin{figure*}[t]
    \centering
    \includegraphics[width=0.8\linewidth]{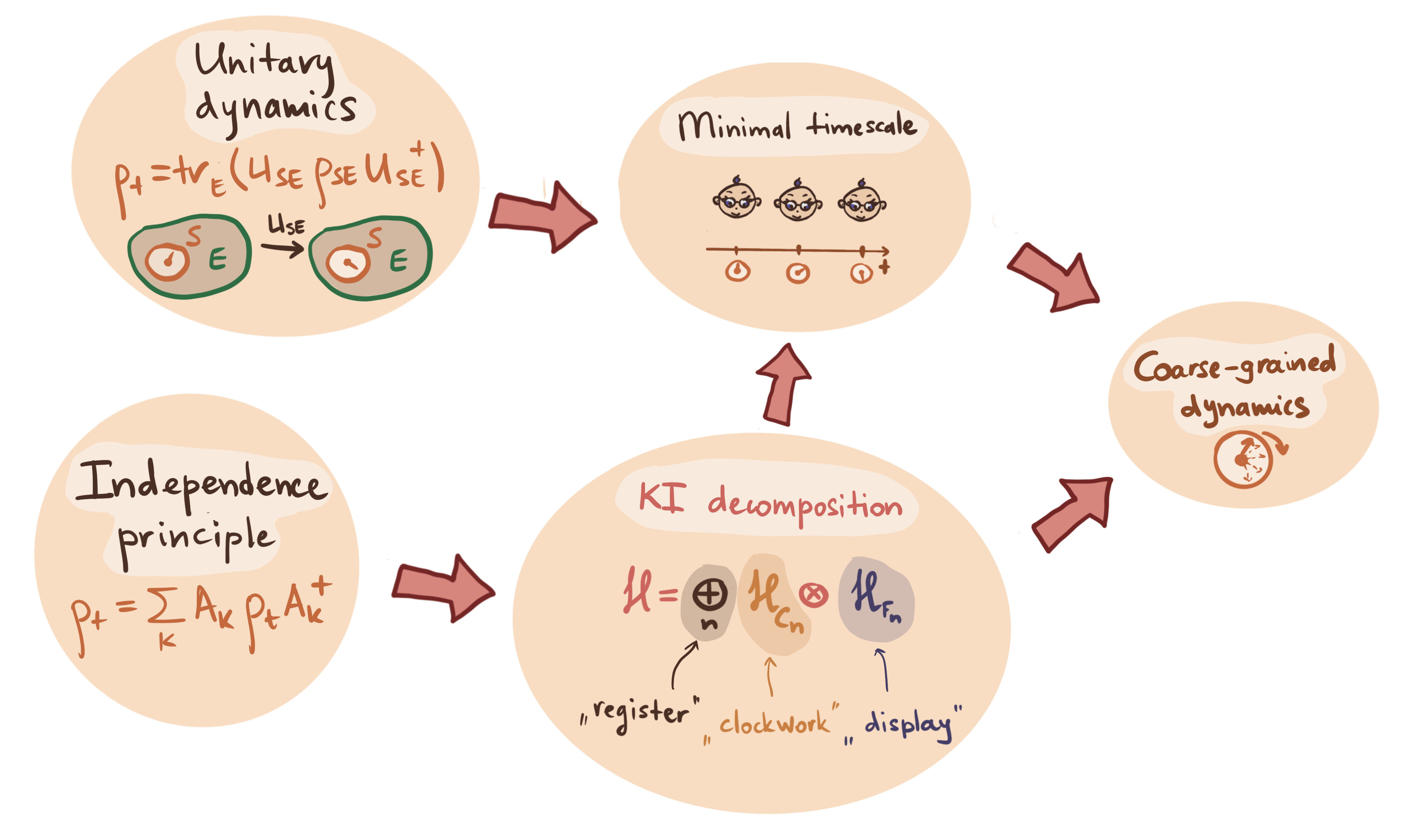}
    \caption{{\bf Structure of a general ticking clock.} The principle of independence implies a Koashi-Imoto decomposition for the clock structure, featuring a classical degree of freedom that stores the time information; to maintain this structure the clock's unitary dynamics have to be coarse grained over, implying a minimal spacing and uncertainty for measurements upon the clock.}
    \label{fig:general-clocks}
\end{figure*}
The passage of time is measured by us via clocks. While the word `clock' usually brings to mind the specialised devices humans have constructed, in a broader sense one can include any system in nature that changes appreciably in time. In fact, animals and plants --- including humans --- employ simpler clocks to regulate their behaviour, such as the circadian rhythm and the solar and lunar cycles.

Starting from the premise that nature is well described by quantum mechanics, we would describe a clock as a system whose state is undergoing some dynamical evolution with respect to background time, and which is occasionally measured by an observer, human or otherwise. However, we cannot usually label as `clock' only the evolving system --- the measurements can disturb the system, and so to understand the behaviour of the clock we must include in the description both the evolving system and the measurements.

This motivates the central theme of this work: how does one get `independent' clocks, i.e. those clocks defined by the property of not being disturbed by the measurements made upon them? This is not only an abstract query: everyday clocks appear to work independently of us observing them. To explain the emergence of such systems from first principles is to explain the theoretical foundation of time-keeping.

In this paper we realize the above in an axiomatic manner. Using existing results from quantum theory, we first derive the general structure of an independent clock, featuring a classical degree of information on which the time information is stored, and an internal inaccessible degree of freedom that may be quantum mechanical. Furthermore, it is argued that these clocks can only arise as a semi-classical limit of a quantum mechanical clock; their evolution is a coarse-grained version of detailed unitary evolution.

With the general form of an independent clock in place we then discuss four natural properties that a clock may possess, each simplifying the description of the clock. Taken all together, they pinpoint what we term an `elementary clock': these include the everyday clocks we are used to, such as wall clocks and wristwatches.

As a consequence, this paper also explains the foundations of `ticking clocks'. This is a recent manner of modelling clocks with the following features: that the clock measures time in discrete steps (ticks) and that these ticks are stored in a classical degree of freedom. Indeed while the original motivation of this work was to establish the theoretical foundations of ticking clocks, our conclusion is that independent clocks are ticking clocks. Specifically: the defining traits of ticking clocks follow naturally by assuming an independent clock.

Finally, we glimpse at the information theory of such clocks: how to characterise the clocks behaviour over time and how this is reflected in measurements upon it. A key point here is that observers in quantum mechanical theory cannot directly measure background time, thus the only way to follow a clocks behaviour is to compare it to other clocks. This leads to the picture of `tick sequences' from multiple clocks --- these are also seen to arise as a specific case of independent clocks.

All in all, this paper serves the following purposes:
\begin{itemize}
	\item to determine the properties and explain the emergence of independent --- i.e. ticking --- clocks,
	\item to classify the various types of ticking clocks, zeroing in upon the properties of the elementary one,
	\item to discuss the foundations for an information theory of these clocks, and finally,
	\item to place recent models of these clocks and results therein into context.
\end{itemize}

The paper is organised as follows. Sec. \ref{sec:generaltickingclock} derives the structure and a constraint on the dynamics of a general ticking clock from the independence principle. In Sec. \ref{sec:elementarytickingclock}, we discuss four further properties that a ticking clock may possess, ending with the model of the elementary ticking clock typically encountered in the literature. Sec. \ref{sec:abstractinfo} discusses the many ways of quantifying the abstract information content of a ticking clock, followed by Sec. \ref{sec:multipleclocks} that discusses the operational scenario of measuring clocks w.r.t. one another. Finally, we wrap up with a discussion and outlook in Sec. \ref{sec:outlook}.

\section{General ticking clocks}\label{sec:generaltickingclock}

\subsection{The notion of a clock and the principle of independence}\label{sec:axiom}

We take a clock to be any object that can be observed in order to make an inference about time, which is assumed to be the continuous background parameter $t$ in the Schrödinger picture of quantum mechanics. Namely,
\begin{itemize}
    \item a clock is a system $S$, whose state depends on $t$;
    \item there exists at least one measurement corresponding to the extraction of time information.
\end{itemize}
The above description includes almost all objects in nature, as it should --- most objects in nature do reflect the passage of time in some fashion and so are in the most basic sense a clock. We now state our central thesis to pick out ticking clocks: the principle of independence. 
\begin{axiom}[Independent clocks]
    A ticking/independent clock is one for which the measurement of time information does not disturb the clock or affect the result of future measurements.\footnote{If our use of the word "ticking" falls in any way afoul of the reader's preferred semantics vis-a-vis clocks, one may replace it with "independent" in the sense implied by the principle, and consider this entire paper to pertain to ``independent clocks''.}
    \label{axiom:independence}
\end{axiom}
While this principle could be applied to a broad range of physical theories, in this paper we focus on clocks that have a quantum mechanical description. The state of the clock is then described by a time-dependent density matrix ${\rho}_t \in L ({\mathcal{H}})$, where ${\mathcal{H}}$ is the Hilbert space of the clock with (bounded) linear operators $L(\mathcal{H})$, and the measurement on the clock by a set of Kraus operators $\{{A}_k\}_k$, with ${A}_k \in L({\mathcal{H}})$ and $\sum_k {A}_k^\dagger {A}_k = {\mathbbm{1}}$~\footnote{The analysis can be generalized to a set of measurements, however a single instance suffices to demonstrate the results.}. 

Furthermore, the dynamics of the clock between measurements is described by the Schr\"odinger equation; this need not be upon the clock alone but upon a larger system of which the clock is a part of. Labelling the rest of the system as the environment $E$, the state of the clock evolves as
\begin{align}\label{eq:detailedevolution}
    \rho_t &= \Tr_E \left[ e^{-i H_{SE} t} \rho_0^{SE} e^{+i H_{SE} t} \right],
\end{align}
where $\rho_{SE}^{(0)}$ is the joint state of the clock and its environment at some time labelled as $t=0$ for simplicity, and $H_{SE}$ is the total Hamiltonian operator.

In the following two subsections we discuss the two consequences of the independence principle: the Koashi-Imoto decomposition of the clock structure and the uncertain timescale of measurements. The logical steps are visualised in Fig.~\ref{fig:general-clocks}.

Before we proceed, note that the principle is only consequential in the quantum regime, as the disturbance of quantum systems is fundamentally unavoidable while measuring. In a regime where classical physics suffices to describe both the clock and the measurement, there is no apparent difference between clocks in general and ticking clocks. We comment further upon this in the discussion.

\subsection{The structure of a ticking clock: emergence of a classical register}\label{sec:structure}

The non-disturbance part of the independence principle is expressed technically as
\begin{align}
	\sum_k {A}_k {\rho}_t {A}_k^\dagger &= {\rho}_t \; \forall t.
\end{align}

A direct consequence of the above is that the clock and the measurement must admit a decomposition given by the Koashi-Imoto theorem~\cite{koashi_operations_2002,hayden_structure_2004,jencova_sufficiency_2006,kuramochi_accessible_2018}. We provide a brief formal introduction to the theorem in Appendix~\ref{appendix:KI}, and state its consequences for clocks below.

The Hilbert space of an independent clock splits into three distinct degrees of freedom, one of which is not accessed by the measurement, the second storing no information vis-\`a-vis $t$ and the the third being classical and accessed in a read-only manner. The first degree of freedom corresponds to a set of `clockwork' spaces $\{\mathcal{H}_{C_n}\}_n$, and the second one to a set of `display' spaces $\{\mathcal{H}_{F_n}\}_n$. We also identify the index $n$ as the third degree of freedom, called the `register' and denoted by $T$. More precisely:
\begin{theorem}[Koashi-Imoto decomposition of a ticking clock]
\label{theorem:KIdecomposition}
The Hilbert space of a ticking clock is of the following form:
\begin{align}
    \mathcal{H} &= \bigoplus_n \mathcal{H}_{C_n} \otimes \mathcal{H}_{F_n},
\end{align}
and the states of the clock $\rho_t$ are block-diagonal w.r.t. the register index $n$:
\begin{align}
    {\rho}_t &= \bigoplus_n p_{n|t} \; \rho_{n|t} \otimes \omega_n,
\end{align}
where
\begin{itemize}
    \item $\omega_n$ are independent of $t$,
    \item $p_{n|t}$ are probability distributions: $p_{n|t} \geq 0$ and $\sum_n p_{n|t} = 1 \; \forall t$,
    \item $\rho_{n|t}$ are normalised states: $\rho_{n|t} \geq 0$ and $\Tr[\rho_{n|t}]=1 \; \forall n$.
\end{itemize}
Furthermore, the measurement channel is subject to the same decomposition: each Kraus operator is of the form
\begin{align}
    {A}_k &= \bigoplus_n \mathds{1}_{\mathcal{C}_n} \otimes A_{k,n}, \\
    \omega_n &= \sum_k A_{k,n} \omega_n A_{k,n}^\dagger \quad \forall n.
\end{align}
\end{theorem}

We see that the clockwork degree of freedom is not accessed by the measurement, while the display degree of freedom holds no temporal information --- the states $\omega_n$ therein are independent of $t$. The measurement only accesses the register index $n$: the probability of obtaining the outcome $k$ from the measurement is
\begin{align}
    P_t(k) &= \sum_n p_{n|t} \Tr \left[ A_{k,n}^\dagger A_{k,n} \omega_n \right] \\
    &= \sum_n p_{n|t} a_{k,n},
\end{align}
where for each $n$ the $\{a_{k,n}\}_k > 0$ are a probability distribution that is independent of time. Any measurement of this form is a \textit{coarse-graining} of the direct measurement of $p_{n|t}$; this is the measurement that simply checks which of the spaces w.r.t. the index $n$ the clock is in, i.e the one with projective Kraus operators:
\begin{align}\label{eq:finegrainedmeasurement}
    {\Pi}_k &= \bigoplus_n \delta_{kn} \mathds{1}_{\mathcal{C}_n}  \otimes \mathds{1}_{\mathcal{F}_n},
\end{align}
for which the probability of the outcome $k$ is simply $P_t(k) = p_{k|t}$.

\begin{figure}
    \centering
    \includegraphics[width=.4\textwidth]{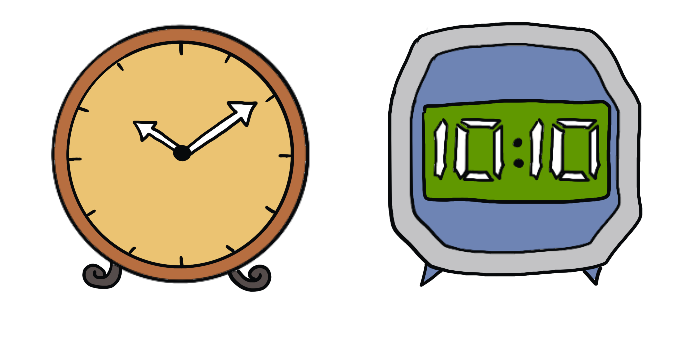}
    \caption{{\bf Two clocks in the same register state.} Even though the display systems are different, one mechanical and the other digital, the time information is equivalent --- 10:10.}
    \label{fig:KIclock}
\end{figure}

To illustrate the roles of the three parts of the clock, take the two clocks as depicted in Fig. \ref{fig:KIclock}, one mechanical and the other digital, both measuring time down to a minute. Within each there is an internal clockwork that is unobserved by us, a mechanical setup in the former and an electronic circuit in the latter. There is also a set of display states for each clock: the different possible configurations of the hands in one case, and the possible states of the digital display in the second case. 

Something that is intuitively clear to us from observation and which the KI decomposition captures elegantly is that the display acts only as an indicator for the time measurement, while not partaking in the actual dynamical time-keeping of the clock. In this example, the time indicated by both clocks is 10:10 --- this corresponds to the index $n$ of the register. If a person wearing one of these clocks is asked what the time is, the answer is 10:10, and not ``10:10 indicated by angles between the hands'' or ``10:10 indicated by electronic ink''.

Thus if one is interested (as we are) in modelling the time-keeping properties of a clock and not the manner in which the information is presented to the environment, then the display part of the clock can be ignored. Mathematically this corresponds to reducing all of the $\mathcal{F}_n$ to trivial spaces, resulting in the following `sufficient' structure for a ticking clock and the measurement upon it:

\begin{thmcorollary}[Minimal ticking clock]
For every ticking clock with Hilbert-space $\tilde{\mathcal H}$, quantum states $ \tilde \rho_t$ and measurement operators $\tilde \Pi_k$ there exists a \emph{minimal clock} $(\mc H,\rho_t,\Pi_k)$ with associated quantum states $\rho_t$ and measurement operators $\Pi_k$ on a Hilbert-space $\mathcal H$ together with completely positive trace-preserving maps $S$ and $R$ fulfilling
\begin{align}
    S( \tilde \Pi_k  \tilde \rho_t)=\Pi_k \rho_t,\ R(\Pi_k \rho_t) =  \tilde \Pi_k  \tilde \rho_t \quad \forall t.
\end{align}
In particular $S( \tilde \rho_t) = \rho_t$ and $R(\rho_t) =  \tilde \rho_t$ for all $t$. Moreover, the Hilbert-space, quantum states and measurement operators of the minimal clock decompose as
\begin{subequations}\label{eq:KIsimple}
\begin{align}
    \mathcal{H} &= \bigoplus_n \mathcal{H}_{C_n}, \\
    \rho_t &= \bigoplus_n p_{n|t} \; \rho_{n|t}, \\
    \Pi_k &= \bigoplus_n \delta_{kn} \; \mathds{1}_n.
\end{align}
\end{subequations}
\end{thmcorollary}
One may wonder how unique the minimal clock. To explore this, let us define two ticking clocks $\rho_t$ and $\sigma_t$ with measurement operators $ \Pi_k$ and $ \Pi'_k$ to be \emph{dynamically equivalent}  (shortened to d-equivalent) if there exist completely positive trace-preserving maps $S$ and $R$ such that
\begin{align}
    S( \Pi_k  \rho_t) =  \Pi'_k  \sigma_t \ R( \Pi'_k  \sigma_t)= \Pi_k \rho_t\quad \forall t.
\end{align}
In particular, a ticking clocks is d-equivalent to its associated minimal clock and two d-equivalent ticking clocks must also have d-equivalent minimal clocks.
In fact, a result about sufficiency in quantum statistical experiments implies that the associated minimal clocks are \emph{unitarily} equivalent \cite{jencova_sufficiency_2006,kuramochi_accessible_2018,galke_sufficiency_2023}:
\begin{thmcorollary}[Dynamical equivalence of clocks]
Two ticking clocks $\tilde \rho_t$ and $\tilde \sigma_t$ with measurement operators $\tilde \Pi_k$ and $\tilde \Pi'_k$ are dynamically equivalent if and only if the associated minimal clocks $(\mc H,\rho_t,\Pi_k)$ and $(\mc H',\sigma_t,\Pi'_k)$ are unitarily equivalent: There exists a unitary operator $U:\mc H\rightarrow \mc H'$ independent of $t$ and such that $U \Pi_k = \Pi'_k U$ and
\begin{align}
U\rho_t U^\dagger = \sigma_t\quad \forall t.
\end{align}
\end{thmcorollary}
Thus two $d$-equivalent clocks essentially correspond to a choice of basis on the Hilbert-spaces for their minimal clocks.

A direct consequence of the d-equivalence of clocks is that $p_{k|t}$ is the same function for all equivalent clocks. In particular, if a clock is equivalent to itself when shifted by some period $\tau$, we find that $p_{k|t}=p_{k|t+\tau}$, which should indeed be the case for any periodic clock.

\subsection{The dynamics of a ticking clock: a minimum uncertain time-scale between measurements}\label{sec:dynamics}

The second part of the independence principle is that the measurement of time information does not affect the result of future statistics. This requires a discussion of the dynamics of the clock in-between measurements. Our starting point is that of standard quantum mechanics: the evolution is described by the Schr\"odinger equation, either upon the clock itself or a larger system that the clock is part of. In the latter case we label the rest of the system as the environment $E$, and represent the evolution of the clock alone as the channel \eqref{eq:detailedevolution}.

For the clock to be non-trivial, $p_{n|t}$ should not be a constant. Consider one such time $t$ at which the probability is changing, so that
\begin{align}
   \frac{\mathrm d}{\mathrm d t} p_{n|t} &= - i \Tr \left[ \left( {\Pi}_n \otimes \mathds{1}_E \right)[H,\rho] \right]\\
    &= -i \Tr\left[[\rho,{\Pi}_n \otimes \mathds{1}_E]H\right]\neq 0.
\end{align}
 Note that although the state of the clock alone is block-diagonal w.r.t. the register index $n$, this need not be true of the joint state $\rho_{SE}$, which is why the above expression can be non-zero.  However, if a time measurement is performed at $t$, then the final state will indeed be decohered w.r.t. $n$ on the full state of clock and environment. But then the expression above becomes zero, and the measurement has disturbed future statistics.

This is not a novel insight, rather it is arises from the same origin as that of the Zeno effect \cite{degasperis_does_1974}. Just as in the Zeno case, here too we have that the evolution of a quantum system from one state to an orthogonal one --- different $n$ spaces --- is disturbed by measurements in that basis; the evolution can be slowed down by measurements, and even frozen in the limit that the measurements are infinitely dense in time.

As a consequence an independent clock in the sense of our principle cannot arise if measurements of time information are unconstrained. The question is therefore, what constraint is necessary to allow for a non-trivial independent clock?

Clearly one necessary condition is that the measurements are spaced out enough to allow for the clock to evolve. But this is not sufficient on its own, one also requires an uncertainty in when the measurements happen. Consider the case that the measurements are spaced out but perfectly timed. Then the only manner in which the future statistics of the clock are not disturbed is if the time period of the measurement is fine-tuned to that of the joint Hamiltonian $H_{SE}$, so that at the moment of measurement there is always zero coherence w.r.t. the register index. This violates the independence principle in spirit as the clock and measurement are constructed to match each other rather than the clock existing as an separate entity; also the measurement itself would require a perfect clock, and one that renders the clock of interest redundant.

We thus arrive at a necessary condition for an independent clock:
\begin{consequence}[Minimal and uncertain measurement timescale]
\label{corollary:min-time}
If one has a clock whose evolution is independent of the extraction of time information, then it must evince a spacing between measurements that is both lower bounded as well as uncertain.
\end{consequence}


Having an uncertainty in the measurement times allows for a \textit{coarse-graining of the dynamics}: within the unitary evolution, there are oscillations of various frequencies, corresponding to (differences of) energy eigenvalues in the full Hamiltonian $H_{SE}$. Those oscillations whose time period is much smaller than that of the uncertainty in the measurement timescale (i.e. of high enough frequency) can be averaged out. The resultant description of the dynamics is not one for infinitesimal $dt$ but rather for $\delta t$ that is of the order of uncertainty in the measurement spacing.

This allows for the emergence of classical degrees of freedom within the clock which is required for the register degree of freedom to exist. Even though the transition between states of the register is effected by underlying unitary dynamics, these could be fast enough w.r.t. the measurement uncertainty so that the coherences between register states average out, leaving it apparently classical. This is a standard manner of describing the \textit{effective} dynamics of open quantum systems \cite{breuer_theory_2007}, and typically requires the environment to be ideal in several senses: infinite, with short correlation timescale, etc.

The conclusion is that ticking clocks are an emergent phenomenon, in much the same way that classical physics at the macroscopic scale is understood to be emergent from microscopic dynamics that is quantum mechanical. In particular, the frequency with which the clock can be measured or interacted with by an observer is constrained.
In a similar way, we think of the Koashi-Imoto decomposition arising from the principle of independence as an idealized emergent property that only holds to very high precision in practice.

\subsection{Discrete ticking clocks vs quantum metrology}

A key feature of ticking clocks is that they appear to measure time in terms of discrete events, hence the name. This arises naturally from the independence principle: the clock structure is such that the only observable change in the clock is when it changes the index of the register. One can thus associate the `tick' to the event of jumping from one state of the register to another.

A direct consequence is that the observed information about time is in terms of a fundamental unit, encoded in the index $n$ of the register. This distinguishes ticking clocks from the general prepare-and-measure scenarios studied extensively in quantum metrology. There too a measurement can be made upon a state $\rho_t$ to estimate the parameter $t$, but without the constraint that the measurement is non-disturbing. As a result, the set of estimated values of $t$ corresponding to all of the possible measurement outcomes need not be discrete, even though each individual measurement typically has a finite --- and thus discrete --- set of outcomes. In appendix~\ref{appendix:discretetick} we illustrate this distinction for the simple quantum clock of Salecker, Wigner and Peres~\cite{salecker_quantum_1958,peres_measurement_1980}.

Thus results from the field of quantum metrology are not automatically applicable to ticking clocks, even though both are heavily concerned with the accuracy of time-keeping. Whether one can still draw a fundamental connection between them remains an interesting open question.

\section{From general to elementary ticking clocks}\label{sec:elementarytickingclock}

\begin{figure}[b]
    \centering
    \includegraphics[width=0.5\textwidth]{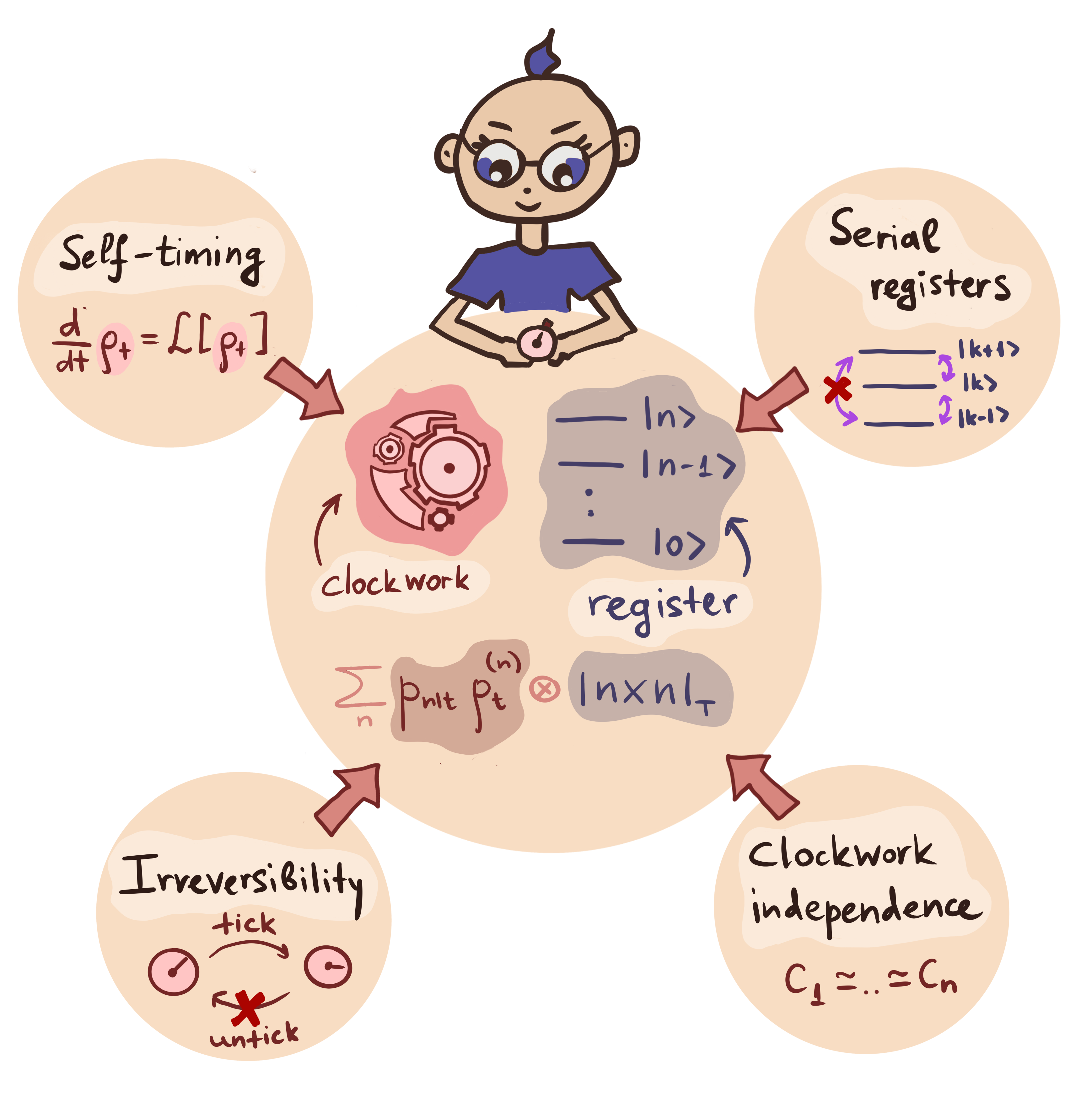}
    \caption{{\bf The assumptions that simplify a general ticking clock to an elementary one.} Listed clockwise from top-left they are: \textit{self-timing}, the requirement that the clock dynamics are self-contained and memory-less --- resulting in Lindbladian-generated dynamics $\frac{\mathrm d}{\mathrm d t} {\rho}_t = {\mathcal{L}} [{\rho}_t]$; \textit{clockwork independence}, the assumption that the dynamics of the clockwork is independent of the register --- resulting in an explicit register system; \textit{serial registers}, the simplification that register states form a chain that the clock only shifts one step upon --- resulting in registers that are tick counters;  \textit{tick irreversibility}, the property that the clock only shifts the register forward --- resulting in the concept of a \textit{time of arrival} for the ticks of a the clock.}
    \label{fig:elementary-clocks}
\end{figure}

We have derived the structure of a general ticking clock together with a restriction on its dynamical description. 
This model differs from the ones typically encountered in the literature~\cite{rankovic_quantum_2015,barato_cost_2016,woods_quantum_2022,woods_autonomous_2021}, which have a simpler structure and regular dynamics. In this section, we describe four additional properties that a clock can possess that lead to a simpler reduced model. We call the clock satisfying all four properties an \textit{elementary} ticking clock.

For each property, we provide the motivation thereof, which types of clocks it excludes, and how the reduction affects the information-theoretic aspects of the clock.

To preview the simplifying properties we state the resulting definition of an elementary ticking clock. The assumptions leading to it are depicted in Fig. \ref{fig:elementary-clocks}.

\begin{definition}[Elementary ticking clock]
\label{definition:elementary}
An elementary ticking clock is a (minimal) ticking clock having the following Hilbert space, state and representative measurement operators:
\begin{subequations}\label{eq:simpledecomposition}
\begin{align}
    {\mathcal{H}} &= \mathcal{H}_C \otimes \mathcal{H}_T, \\
    {\rho}_t &= \sum_n p_{n|t} \; \rho_{n|t} \otimes \ket{n}\!\bra{n}_T, \\
    {A}_k &= \mathds{1}_C \otimes \ket{n}\!\bra{n}_T,
\end{align}
\end{subequations}
and whose dynamics is governed by a Lindbladian master equation of the form
\begin{align}\label{eq:elementaryLindbladian}
    \frac{\mathrm d}{\mathrm d t} {\rho}_t = {\mathcal{L}}[\rho_t] &= -i \left[ H_C \otimes \mathds{1}_T,\rho_t \right] + \sum_k \mathcal{D}_{L_k \otimes \mathds{1}_T} [\rho_t] \nonumber \\
    &\quad\quad\quad\quad\quad\quad\;\;+ \sum_j \mathcal{D}_{J_j \otimes \Gamma_T} \left[ \rho_t \right].
\end{align}
Here, $\mathcal{D}_X [\rho]$ is the standard dissipator $\mathcal{D}_X [\rho] = X \rho X^\dagger - \frac{1}{2} \left\{ X^\dagger X, \rho \right\}$ and $\Gamma$ is the shift operator on the register $\Gamma = \sum_n \ket{n+1}\!\bra{n}_T$.
\end{definition}

\subsection{Self-timing}

A general dynamical description of a clock requires one to include and keep track of the degrees of freedom within the environment that interact with the clock. However, in investigating the fundamental nature of ticking clocks one is typically interested in how the temporal behaviour of the register is determined by the properties of the clock only -- which is impossible to conclude if the said behaviour derives from some external timing, whether it is realized via significant correlations with environment, or via the time-dependence of the dynamics of the clock. If we assume that the dynamics of the clock are Markovian (memoryless), and do not require keeping track of the environment, we arrive to a simpler description of the clock's dynamics. We call this pair of assumptions \textit{self-timing} in the same manner as Ref.~\cite{woods_autonomous_2021}.

\begin{property}[Self-timing]
\label{property:self-timing}
A clock is \textit{self-timed} if
\begin{enumerate}
    \item it does not require to keep track of the environment: if the state at some time $t=0$ is ${\rho}_0$, the state at any other time $t>0$ depends only upon the clock itself via
    \begin{align}
        {\rho}_t &= {\Lambda}_t [{\rho}_0],
    \end{align}
    where $\{ \Lambda_t\}$ is a family of completely positive trace preserving maps;

    \item its dynamics are Markovian: the maps $ \Lambda_t$ are CP-divisible, i.e. for all $s,t$ such that $0 \leq s \leq t$,
    \begin{align}
        {\Lambda}_t &= {\Lambda}_s \circ {\Lambda}_{t-s}.
    \end{align}
\end{enumerate}
\end{property}
Together these imply that the set of maps $\{ \Lambda_t\}_t$ form a dynamical semi-group~\cite{lindblad_generators_1976} whose generator is labelled a \textit{Lindbladian} operator ${\mathcal{L}}$:
\begin{align}
    {\Lambda}_t &= e^{{\mathcal{L}} t}, \\
    \frac{\mathrm d}{\mathrm d t} {\rho}_t &= {\mathcal{L}} [{\rho}_t].
\end{align}

However, not every Lindbladian generator is suitable to describe a ticking clock: the only allowed dynamics are those which preserve the algebra of the KI decomposition of the clock's state~\eqref{eq:KIsimple}. Employing a recent result~\cite{hasenohrl_generators_2022}, which provides the description of generators that have an invariant atomic algebra, we arrive to the following characterization of their dynamics.
\begin{theorem}[Dynamics of a self-timed ticking clock]\label{theorem:self-timing}
For a minimal ticking clock whose dynamics is described by a semi-group, the generator is of the form
\begin{subequations}
\begin{align}\label{eq:firstLindblad}
    {\mathcal{L}} \left[ {\rho}_t \right] &= - i \left[ {H}, {\rho}_t \right] + \sum_j \gamma_j \mathcal{D}_{{L}_j} \left[ {\rho}_t \right],
\end{align}
where the Hamiltonian is block-diagonal w.r.t. the register index $n$:
\begin{align}
    {H} &= \bigoplus_n H^{(n)}_{C_n},
\end{align}
and the `jump operators' ${L}_j$ may only connect a single clockwork space to a single other one (possibly the same), i.e. for every ${L}_j$ there exists only one pair of register indices $\{m,n\}$ (possibly equal) such that
\begin{align}\label{eq:singularjump}
    {L}_j &= \Pi_m {L}_j \Pi_n,
\end{align}
\end{subequations}
where $\Pi_m = \bigoplus_k \delta_{km} \mathds{1}_{\mathcal{C}_k}$ is the projector upon the $m^{th}$ subspace of the clockwork.
\end{theorem}
We see that the dynamics of the clock decomposes into two parts: one corresponding to general Lindbladian dynamics within each clockwork space --- encoded in the block-diagonal Hamiltonian ${H}$ and the jump operators ${L}_j$ that keep the clock in the same clockwork space --- and the other corresponding to jumps between clockwork spaces. Both sets of operators leave the register classical, i.e. they introduce no coherence between clockwork subspaces.

Self-timing is a strong condition and therefore self-timed ticking clocks exclude some examples of ticking clocks, for instance: clocks driven by externally timed signals; clocks that amplify the accuracy of an incoming temporal signal; and clocks for which one wishes to describe only a sub-part of the entirety of the clock. For an example of all three see~\cite{yang_accuracy_2019}.

In a very strict sense, self-timing is not even fulfilled by the best clocks that we use today, namely atomic clocks. These rely on synchronizing the frequency of an oscillator (a laser or maser) to atomic transition frequencies. While the bare atomic transition frequencies are in principle fixed and an \emph{ideal} atomic clock may be considered self-timed, in reality the frequency of the oscillator drifts in time due to environmental influences.

Therefore, one should view self-timing as an idealisation that is fulfilled up to some precision, in the same way as the KI decomposition.

\subsection{Clockwork independence}

In the general model~\eqref{eq:KIsimple}, we distinguish between distinct clockwork spaces for each $n$, corresponding to distinct dynamics for each state of the register. This is useful for descriptions of multi-stage processes that can be interpreted to be a ticking clock by associating the label of the stage to the register index. The dynamics of each stage is in general distinct, and their modelling would require distinct clockwork spaces. A decay chain of radioactive isotopes is one example of such a process, as is protein synthesis within organisms.

It is no coincidence that the above examples feature only a small number of stages --- a clock that requires a separate clockwork for each value of the register index grows in size as the register size gets larger. A useful property for a ticking clock is thus that of \textit{clockwork independence}: that the dynamics of the clockwork is independent of the state of the register. This property allows for the same clockwork to be used throughout, which in turn allows for larger registers, possibly even of an unbounded size.

\begin{property}[Clockwork independence]
\label{property:clockwork-independence}
    The clockworks of a ticking clock are independent of the state of the register if
    its description can be simplified to a separate Hilbert space and state for the register, taking us from the general minimal clock \eqref{eq:KIsimple} to that of the simple model \eqref{eq:simpledecomposition}:
    \begin{subequations}
    \begin{align}
        \mathcal{H} &= \bigoplus_n \mathcal{H}_{C_n} & &\longrightarrow \mathcal{H}_C \otimes \mathcal{H}_T, \\
        \rho_t &= \bigoplus_n p_{n|t} \; \rho_{n|t} & &\longrightarrow \sum_n p_{n|t} \; \rho_{n|t} \otimes \ket{n}\!\bra{n}_T, \\
        A_k &= \bigoplus_n \delta_{kn} \; \mathds{1}_n & &\longrightarrow \mathds{1}_C \otimes \ket{n}\!\bra{n}_T,
    \end{align}
    \end{subequations}
    with the dynamics simplifying analogously, the Hamiltonian in~\eqref{eq:firstLindblad} now being the tensor product
    \begin{align}
    {H} &= H_C \otimes \mathds{1}_T,
    \end{align}
    and the Lindblad jump operators splitting similarly:
    \begin{align}\label{eq:indLindbladjump}
        {L}_{k,\Delta} &= L_{k|\Delta} \otimes \Gamma^{\Delta},
    \end{align}
    where each $\Gamma^{\Delta}$ is a translation by $\Delta \in \mathbb{Z}$ steps of the register
    \begin{align}
        \Gamma^{\Delta} &= \sum_n \ket{n + \Delta}\!\bra{n}.
    \end{align}
\end{property}

With the assumptions of self-timing and clockwork independence we gain a significant feature of the model: the stabilisation of the clock dynamics over large times (more precisely, the property of asymptotic stationarity). This is discussed further in Sec. \ref{sec:abstractinfo}.

Clockwork independence also appears as an axiom in \cite{woods_autonomous_2021}, where it is termed the `time invariance symmetry condition'.

\subsection{Serial registers}

In our general model, the dynamics of the clock can involve jumps between any pair of states of the register. An example of such a process in nature is a multi-stage chemical reaction: one may skip a stage via an enzymatic shortcut. Here too by identifying the stage with the register index we get a ticking clock with jumps that are not just on sequential states of the register. We will also see that non-serial registers are important for the theory of multiple clocks and tick sequences, Sec. \ref{sec:multipleclocks}.

The clocks that we are used to do typically have \textit{serial registers}, i.e. the states of the register are arranged in a totally ordered chain with jumps only possible between nearest neighbours.
\begin{property}[Serial registers]
\label{property:serial-registers}
    A ticking clock has serial registers if its Lindblad jump operators in \eqref{eq:singularjump}, \eqref{eq:indLindbladjump} are non-zero for at most three values $m-n = \Delta \in \{-1,0,+1\}$.
\end{property}
This assumption also naturally strengthens the interpretation of the register index as the `number of ticks' of the clock.

This property is closely related to the `leading order condition' axiom in \cite{woods_autonomous_2021}.

\begin{table*}[t]
    \centering
    \begin{tabular}{ |c|c|c|c|c| }
\hline
& \textbf{Self-timing} & \textbf{Clockwork independence} & \textbf{Serial registers} & \textbf{Irrev. ticks} \\
\hline
Elementary ticking clock (Def.~\ref{definition:elementary}) & \cmark & \cmark & \cmark & \cmark \\
\hline
Clocks driven by externally timed signals (e.g.~\cite{yang_accuracy_2019}) & \xmark & \cmark & \cmark & \cmark \\
\hline
Atomic clocks & \xmark & \cmark & \cmark & \cmark \\
\hline
 Brownian clock of~\cite{barato_cost_2016}  & \cmark & \cmark &  \cmark & \xmark \\
\hline
Circadian rhythm (see e.g. \cite{foster_2022_life}) & \xmark & \xmark & \xmark & \cmark \\
\hline
\end{tabular}
    \caption{{\bf Different kinds of clocks and their properties.}}
    \label{table:properties}
\end{table*}

\subsection{Irreversible ticks}

The final simplification to consider is that of the irreversibility of the register dynamics: that the clock dynamics can only change the register index in a particular direction. For serial registers, as the index $n$ is associated to the number of ticks, this property is equivalent to the statement that once a tick is recorded it can never be reversed. One may also have irreversibility for non-serial registers as along as a partial order is in place.
\begin{property}
\label{property:irreversibility}
A ticking clock is irreversible if there exists a partial order $\leq$ on the set of register indices $\mathcal I$ such that if the register index is measured to be $n$ at any time $t$, then for all $t^\prime > t$ the only possible outcome of a time measurement is $m \geq n$, where $m,n\in\mathcal I$.
\end{property}

This property appears to matches our experience: every clock that we are used to employing from the pendulum clock down to atomic clocks have ticks that once recorded cannot be 'unrecorded' --- the tick is recorded into the environment in a manner that does not allow for reversal. This has the same flavour as the irreversibility of measurements in quantum mechanics --- from the viewpoint of an observer, the measurement process is irreversible and the outcome cannot be erased~\footnote{Celebrated thought experiments regarding the internal inconsistency of quantum mechanics always involve toying with the reversibility (or lack thereof) of the measurement.}.

The observational irreversibility goes hand-in-hand with thermodynamic irreversibility. Clocks that tick ``loudly'' --- i.e. whose ticks are recorded on a macroscopic scale --- dissipate entropy in their environment as a result. This is because the process of recording a tick on a macroscopic scale involves the production of a macroscopic amount of entropy~\cite{bennett_notes_2003}. 
The reversal of these processes would involve a large decrease in entropy that is in practice not observed. 
In other words the reversal of the tick is possible but improbable, and absent in all practical cases.

This outlook also fits with our understanding of the thermodynamic theory of fluctuations and detailed balance~\cite{evans_probability_1993}. 
Given any dynamical process, there exists its reverse process, and the rates of the two are related to each other by the difference in entropy production in either direction. 
Thus to eliminate the possibility of the reverse of a tick one must have the tick-process create a large amount of entropy in the environment.

All of this suggests the assumption of irreversibility is judicious, with one exception: in the case one would like to understand the \textit{fundamental entropic cost} of a ticking clock. While not currently a question relevant to practical clocks it is still of interest from the point of view of fundamental physics, as it concerns the ``arrow of time'': the observation that entropy increases as we observe time passing. For clocks this turns into a quantifiable question: Does the precision of a clock come at a fundamental entropic cost \cite{erker_autonomous_2017}?

One thus has a choice: if entropic considerations are not central then assuming irreversible ticks is appropriate; this takes us to the model of the elementary ticking clock as described in Def. \ref{definition:elementary}. 
This is also the model employed by~\cite{rankovic_quantum_2015,woods_quantum_2022,woods_autonomous_2021,erker_autonomous_2017,schwarzhans_autonomous_2021}. 
On the other hand, if entropy is the fundamental resource (either directly or in the form of heat or other thermal processes that keep the clock out of equilibrium) then it makes sense to keep the ticks reversible, as is done in \cite{barato_thermodynamic_2015,barato_cost_2016}.

It is also important to understand a finer distinction between thermodynamic reversibility of the ticks, and thermodynamic reversibility within the clockwork: it is quite possible to keep reversibility for processes within the clockwork but not for the ticks. For instance, one may demand that detailed balance is obeyed by the jump operators that do not change the register state, while restricting those that change it to only go forward. This middle ground is the form used in~\cite{erker_autonomous_2017,schwarzhans_autonomous_2021}. It is justified if the entropic dissipation within the clockwork is high enough as to render the ignored entropic cost of the backward tick insignificant; the conclusions drawn in such models are hence applicable in the high dissipation regime.

The major advantage in having irreversible ticks is that there is now a single (but random) time that the register arrives to an index $n$. This \textit{time of arrival of a tick} is an alternative description of the register dynamics to that of the \textit{number of ticks}. We discuss this in greater detail in the next section.

All of the properties discussed in the section can be considered independently of each other; in Table~\ref{table:properties} we list examples of different clocks and indicate which of the four properties they satisfy.

\section{The information content of a ticking clock}\label{sec:abstractinfo}

In this section, we lay out the basic principles of an information theory of ticking clocks; namely, we identify the objects encoding their information content, and connect various measures of the accuracy of ticking clocks. The full treatment of the information theory is beyond the scope of this paper, and will be addressed in future work~\cite{RefClock,InfoClock}.

How do we extract the information from a ticking clock? The first thing to note is that the only way we access the clock's information content is via reading off the state of its register. Thus, the information theory of a clock is, strictly speaking, the information theory of the clock's register~\cite{rankovic_quantum_2015,woods_quantum_2022}, and the question of how the properties of the clockwork affect the behaviour of the ticks~\cite{woods_quantum_2022,erker_autonomous_2017,yang_ultimate_2020} could only be answered by tracking the behaviour of the register.

For each time $t$, a measurement of the register will return the answer $n$ with probability $p_{n|t}$. If one performs many such measurements at different times, there is an associated joint probability distribution $p_{n_1,n_2,...n_m|t_1,t_2,...,t_m}$. This joint distribution encodes the information content of a single clock.

For ticking clocks which satisfy the properties of self-timing (Property~\ref{property:self-timing}) and clockwork independence (Property~\ref{property:clockwork-independence}), the theory simplifies considerably. The dynamics are given by a time-independent Lindbladian generator that is translation invariant w.r.t. the register. One can split the Lindblad operator~\eqref{eq:firstLindblad} w.r.t. how much each term shifts the register:
\begin{subequations}\label{eq:fullcountingLindblad}
\begin{align}
    {\mathcal{L}} \left[ {\rho}_t \right] &= \sum_\Delta \left( \mathcal{L}_\Delta \otimes X_\Delta \right) \left[ {\rho}_t \right], \\
    \quad X_\Delta \big[ \ket{n}\!\bra{n} \big] &= \ket{n+\Delta}\!\bra{n+\Delta} \; \forall n.
\end{align}
For each $\Delta \neq 0$ the $\mathcal{L}_\Delta$ is an unnormalised channel
\begin{align}
    \mathcal{L}_\Delta [\rho] &= \sum_{k} \gamma_{k,\Delta} L_{k|\Delta} \rho L_{k|\Delta}^\dagger,
\end{align}
while the $\Delta=0$ operator collects all terms in the generator that leave the register unchanged ($
X_0 = \mathds{1}$),
\begin{align}
    \mathcal{L}_0 \left[ \rho \right] &= -i \left[ H, \rho \right] - \frac{1}{2} \sum_{k,\Delta} \gamma_{k,\Delta} \left\{ L_{k|\Delta}^\dagger L_{k|\Delta}, \rho \right\}.
\end{align}
\end{subequations}
It is customary to refer to $\mathcal{L}_0$ as a non-tick generator, and $\mathcal{L}_\Delta$ as ticking generators \cite{woods_quantum_2022}, especially in the case of an elementary clock, for which there is only $\mathcal{L}_0,\mathcal{L}_1$. One could correspondingly view the process of each tick of the clock as comprised of two processes working in tandem: the internal evolution of the clockwork generated by $\mathcal{L}_0$ (labelled \textit{temporal probability concentration}~\cite{schwarzhans_autonomous_2021}) and the tick process generated by $\mathcal{L}_1$.

The set $\{\mathcal{L}_\Delta\}_\Delta$ allow for the description of the clock without the need of an explicit register, by following the evolution of the set of clockwork states:
\begin{align}
     \frac{\mathrm d}{\mathrm d t} \left( p_{n|t} \rho_{n|t} \right) &= \sum_\Delta \mathcal{L}_\Delta \left[ p_{n-\Delta|t} \rho_{n-\Delta|t} \right].
\end{align}

\subsection{Asymptotic stationarity and full counting statistics.} 
\label{subsec:counting-statistics}
In general, in order to capture the full information content of a ticking clock which runs indefinitely, one has to consider the joint probability $p_{n_1,n_2,...n_m|t_1,t_2,...,t_m}$ for an infinite range of times $t_1$ to $t_m$ and an infinite density of measurements in between. However, this is rendered unnecessary for self-timed clocks with an independent clockwork: as the dynamics are generated by a time-independent Lindbladian the behaviour of the clock in the long-time limit is typically independent of the initial state of clock and register, deriving only from the properties of the generator\footnote{A sufficient condition is that the reduced Linbladian upon the clockwork has a unique steady state}. This ergodic behavior is a well-known concept in the theory of Markov processes. Self-timed clocks in particular fall within the theory of \textit{full counting statistics}~\cite{schaller_technical_2014}~\footnote{To be precise, the consideration of~\cite{schaller_technical_2014} also assumes the clock having serial registers, as it splits the relevant Lindblad clock dynamics into $\mathcal{L}_+$ and $\mathcal{L}_-$, corresponding to increasing and decreasing the tick count of the clock's register. One can however also carry out a similar analysis without demanding that property, with $e^{i*j*\rchi}$ being the appropriate phase field for $\mathcal{L}_j$ terms in~\eqref{eq:fullcountingLindblad}.}. We summarize the main points of the theory below.

Given the probabilities $p_{n|t}$ one can define the moments $\braket{n^k}_t = \sum_n n^k p_{n|t}$. For large $t$, these approach polynomials in $t$, the degree of which is the degree of the moment.
\begin{align}
    \braket{n^k}_t &= \sum_{j=0}^k a_{k,j} t^j.
\end{align}
Every coefficient $a_{k,j}$ with the exception of the constant term $a_{k,0}$ is independent of the initial state of the clockwork and register, and determined solely by the set of Lindbladian operators $\{\mathcal{L}_\Delta\}_\Delta$. The asymptotic behaviour of the clock is thus encoded in these moments.

In particular, the mean and variance of the number of ticks scales linearly for large values of $t$
\begin{subequations}\label{eq:asymptoticrates}
\begin{align}
    \braket{n}_t &\rightarrow \nu \; t, \\
    \braket{n^2}_t - \braket{n}^2_t &\rightarrow \Sigma \; t,
\end{align}
\end{subequations}
where $\nu,\Sigma$ are constants determined by $\{\mathcal{L}_\Delta\}$. The mean and variance are in fact the first and second \emph{cumulants} of $p_{n|t}$, every one of these are asymptotically linear, and the set of them contains the same information as the moments.~\cite{schaller_technical_2014}

\subsection{Irreversible ticks: time of arrival and waiting time.} 

If the clock also satisfies the properties of having serial registers and irreversible ticks (Properties~\ref{property:serial-registers} and~\ref{property:irreversibility}), in other words, it is an elementary clock by Def.~\ref{definition:elementary}, one can also characterize it via a \textit{time of arrival} for the $n^{th}$ tick of the clock, denoted by $T_n$. This denotes the moment of time at which the state of the register changes from $\ket{n-1}_T$ to $\ket{n}_T$. Unless the clock is perfect, the time of arrival does not have a definite value, but is distributed according to the probability density $P(T_n = t)$. The time-of-arrival and number-of-ticks pictures are connected by the following transformation
\begin{align}
    P(T_n = t) &= - \frac{\mathrm d}{\mathrm d t} \sum_{m<n} p_{m|t}.
\end{align}

Just as in the case of the number of ticks, the full information of the clock is not encoded in a single time of arrival, but rather in a joint probability density $P(T_1 = t_1,T_2 = t_2, ..., T_n = t_n)$. In the literature, the time of arrival $T_n$ is referred to as a \textit{waiting time} for the $n^{th}$ tick, and its distribution as a \textit{waiting-time distribution}, also sometimes referred to as a \textit{delay function}~\cite{erker_autonomous_2017,woods_quantum_2022}.

\subsection{Reset clocks: i.i.d. ticks.} 

If the clock has the further simplifying property that it goes to the same fixed state immediately after every tick (we call such clocks \textit{reset clocks}), then the behaviour of each tick becomes identical and independent of the ticks before. The waiting-time distribution of a single tick is hence sufficient to describe everything about the clock.

If we denote this distribution as $\omega_t$, then the joint waiting-time distribution of any number of consecutive ticks is simply the product,
\begin{align}
    P(T_1 = t_1, ..., T_n = t_n) &= \omega_{t_1} \cdot\omega_{t_2-t_1}\cdot ... \cdot \omega_{t_n-t_{n-1}},
\end{align}
and the distribution for the waiting-time of solely the $n^{th}$ tick is the n-fold convolution of $\omega_t$.

\subsection{Precision of ticks.} 

There are several ways to quantify how good a clock is; the simplest one is based on comparing the mean value of some clock quantity to the fluctuations of the same~\cite{barato_cost_2016,erker_autonomous_2017,woods_quantum_2022,yang_ultimate_2020,yang_accuracy_2019,meier_fundamental_2023}. Both the number of ticks $n_t$ and the time of arrival $T_n$ can be chosen the quantity.

The number-of-ticks picture is more general, as it does not require the clock to have serial registers or irreversible ticks. As the asymptotic rates of the mean and the variance approach constant values \eqref{eq:asymptoticrates}, the ratio between said rates is a valid measure of precision:
\begin{align}
    R_1 &= \lim_{t \to \infty} \frac{\braket{n}_t}{\braket{n^2}_t - \braket{n}^2_t} = \frac{\nu}{\Sigma}.
\end{align}
This is the inverse of the `Fano factor' and is the measure employed in~\cite{barato_thermodynamic_2015,barato_cost_2016}, and widely used in the literature on thermodynamic uncertainty relations~\cite{horowitz_thermodynamic_2020}.

An elegant property of the above is its time-scale invariance --- if we transform $t \to t/a$, or equivalently multiply the dynamical generator $\mathcal{L}$ by $a$, the precision remains unchanged. As such it measures the fluctuations of the ticks of the clock w.r.t. the clock's own timescale.

On the other hand, consider that we have an elementary reset clock. Here too there is an analogous time-scale invariant measure of precision:
\begin{align}
    R_2 &= \frac{\mu^2}{\sigma^2},
\end{align}
where $\mu,\sigma^2$ are the mean and variance of the waiting-time distribution $\omega_t$. This is the measure of precision referred to in~\cite{erker_autonomous_2017,woods_quantum_2022,yang_accuracy_2019,yang_ultimate_2020,schwarzhans_autonomous_2021,meier_fundamental_2023}, which corresponds to the number of ticks seen from the clock before the uncertainty in the interval between successive ticks becomes as large as the mean time between them.

As expected, these two precision measures are equal $R_1 = R_2$ (see Appendix~\ref{app:cumulantsvswaitingtime} for a derivation), in particular:
\begin{align}
    \nu &= \frac{1}{\mu}, \quad \Sigma = \frac{\sigma^2}{\mu^3}.
\end{align}

Another standard of precision is the Allan variance, a measure of the stability of the frequency of the clock.  When applied to the case of self-timed clocks operating in the stationary (asymptotic) limit the Allan variance takes a simple form:
\begin{align}
    A_\tau &= 
    \frac{\Sigma}{\tau} = \frac{\nu}{R_1\tau},
\end{align}
where $\tau$ is the averaging time (for a derivation, see Appendix~\ref{appendix:allan-variance}).
Thus for self-timed clocks, the Allan variance does not contain more information than the usual precision $R_1$ and decreases with $\tau$. It becomes an important measure in its own right in the case of non-self-timed clocks, in particular those that exhibit a frequency drift. In this case, the Allan variance for larger values of $\tau$ reflects the frequency drift.

\subsection{Discretisation of dynamics: maps-based picture of ticking clocks.} 

Another way to view the dynamics of the clock is to discretise it into many applications of a family of dynamical maps. This approach represented one of the earlier efforts to model clocks from a quantum information perspective~\cite{rankovic_quantum_2015}, and proved to be useful in deriving fundamental bounds on the precision of clocks \cite{yang_ultimate_2020}. Here we describe it from the perspective of our work.

By taking the evolution of an elementary clock for a small but finite amount of time $\delta$ one can construct a quantum operation as follows. Consider the state of the clock at some time $t$ conditioned on the register being in the state $n$:
\begin{align}
    \rho_t &= \rho_{n|t} \otimes \ket{n}\!\bra{n}
\end{align}
After a time $\delta$, the above transforms to
\begin{align}
    {\rho}_{t+\delta} &= \sigma^{(0)}_{t+\delta} \otimes \ket{n}\!\bra{n} + \sum_{m=1}^\infty \sigma^{(m)}_{t+\delta} \otimes \ket{n+m}\!\bra{n+m},
\end{align}
where the $\sigma_{t+\delta}^{(m)}$ are sub-normalized density matrices. The state of the clockwork alone is
\begin{align}
    \rho_{t+\delta} &= \sigma^{(0)}_{t+\delta} + \sum_{m=1}^\infty \sigma^{(m)}_{t+\delta} \\
        &= \mathcal{M}^{(0)}_\delta [\rho_t] + \mathcal{M}^{(1)}_\delta [\rho_t],
\end{align}
where we have collected all of the states corresponding to one or more ticks having being generated in the second term. The separation above defines two operators, by expanding the evolution operator $e^{{\mathcal{L}} \delta}$ for small $\delta$ these are seen to be of the form:
\begin{align}
    \mathcal{M}^{(0)}_\delta &= \mathds{1} + \delta \mathcal{L}_0 + O \left( \delta^2 \right), \\
    \mathcal{M}^{(1)}_\delta &= \delta \mathcal{L}_+ + O \left( \delta^2 \right),
\end{align}
where $\mathcal{L}_0,\mathcal{L}_1$ are the clock operators corresponding to not shifting and shifting the register respectively, \eqref{eq:fullcountingLindblad}.

This abstract division between the two parts of the clock state can be made into a full quantum operation by including a bit register that indicates whether the clock ticked or not in the time interval $\delta$:
\begin{align}\label{eq:discretemap}
    \mathcal{M}_{C \rightarrow CR} [\rho] &= \mathcal{M}^{(0)}_\delta [\rho] \otimes \ket{0}\!\bra{0}_R + \mathcal{M}^{(1)}_\delta [\rho] \otimes \ket{1}\!\bra{1}_R.
\end{align}
One may now repeat the above map, each time with a new register $R_i$, resulting in a sequence of bits $\{R_i\}_i$ that encode the clock's ticking probability in small time intervals~\cite{rankovic_quantum_2015,woods_quantum_2022}.

As has been noted before~\cite{woods_autonomous_2021}, this picture is not a constructive description: 
in order to apply each map one needs to disconnect register $R_i$ after an interval of precisely $\delta$ and connect $R_{i+1}$, requiring the presence of a precise clock of high frequency in the background, in addition to a high number of bit registers and connecting machinery. In the limit $\delta \rightarrow 0$ --- which is required for the state of the sequence of $R$'s to exactly match the waiting-time distribution --- both the precision of the background clock and the amount of memory become infinite.

Furthermore, recall that a ticking clock can only emerge in a regime where the measurements on the register are both spaced out and uncertain enough in time. Thus, while $p_{n|t}$ and all associated objects are correct predictors of the observable probabilities, they rely on the assumption that one is not measuring frequently enough to invalidate the coarse-grained dynamics. With a high enough density of measurements on the clock one would invalidate the principle of independence, invalidating most of the analysis so far.

However, as an abstract representation of the information content of the clock the model works well. The probability distribution over the sequence of registers is a discretisation (to step size $\delta$) of the distribution of waiting-times. In fact, it can be shown \cite{RefClock} that for $\delta$ small enough (but still non-zero) the discrete statistics are in on-to-one correspondence to those in the continuous picture. It is also worth noting that by starting from discrete maps one can derive that the clock must have Lindbladian dynamics as described in~\cite{woods_quantum_2022}.

\subsection{Abstract vs operational information.} 

All of the quantities of the clock we discuss so far in this section were defined w.r.t. the background time $t$: $p_{n|t},P(T_n=t),\braket{n}_t,\omega_t$, and so on. However, in realistic scenarios we don't have access to perfect background time, but rather infer it from measurements on clocks. In other words, in a strictly operational sense the above quantities are not in fact measurable, as we can only measure quantities related to comparing the ticks of different clocks to each other~\cite{rankovic_quantum_2015}.

The latter is the focus of the next section. 
For now we distinguish between what can be called the \textit{abstract information content} of a single clock, encompassing any such quantity defined w.r.t. background time $t$ and the \textit{operational information content}, which is accessible to us when we have multiple clocks to compare ticks from.

\section{Multiple clocks: tick sequences}\label{sec:multipleclocks}

When we have multiple clocks, we have access not only to the individual ticks, but also to their relative ordering; as a result, the main object considered is the tick sequence capturing that ordering. This is however only applicable to irreversible clocks, as it requires the notion of `recording a tick' to be well-defined.

To illustrate this, let us take two elementary clocks $A$ and $B$ and compose them~\footnote{The model presented in this section can be analogously extended to an arbitrary number of considered clocks.}. Their joint state belongs to a Hilbert space consisting of four subsystems, namely, two clockworks and two corresponding registers. The dynamics can then be separated into two parts
\begin{itemize}
    \item \textit{the internal dynamics}: the dynamics for each clockwork that do not shift the register, encoded by the Hamiltonian $H_{C_A} \otimes \mathds{1}_{\text{rest}}$ and set of jump operators $\{L_{k|0,C_A} \otimes \mathds{1}_{T_A} \otimes \mathds{1}_{\text{rest}}\}_k$ (analogously for $B$),
    \item \textit{the ticking dynamics} of each clock, encoded by the set of operators$\{L_{j|1,C_A} \otimes \Gamma_{T_A} \otimes \mathds{1}_{\text{rest}}\}_j$ (analogously for $B$).
\end{itemize}
We denote the first part (which does not concern the register) by $\mathcal{L}_{0,A} [\rho_t]$ (similarly for $B$), splitting the total Lindbladian as follows:

\onecolumngrid
\hrulefill
\begin{align}
    \frac{\mathrm d}{\mathrm d t} \rho_t &= -i \left[H_{C_A} \otimes \mathds{1}_{T_AC_BT_B}, \rho_t \right] + \sum_k \mathcal{D}_{L_{k|0,C_A} \otimes \mathds{1}_{T_AC_BT_B}} [\rho_t] + \sum_j \mathcal{D}_{L_{j|1,C_A} \otimes \Gamma_{T_A} \otimes \mathds{1}_{C_BT_B}} [\rho_t] + \text{similar for $B$.} \\
    &= \mathcal{L}_{0,A} [\rho_t] + \mathcal{L}_{0,B} [\rho_t] + \sum_j \mathcal{D}_{L_{j|1,C_A} \otimes \Gamma_{T_A} \otimes \mathds{1}_{C_BT_B}} [\rho_t] + \sum_j \mathcal{D}_{\mathds{1}_{C_AT_A} \otimes {L}_{j|1,C_B} \otimes \Gamma_{T_B}} [\rho_t].
\end{align}
\hrulefill
\twocolumngrid

To incorporate the ordering between ticks, we replace the separate registers for $A$ and $B$ by a single register for both that will store the tick sequence. Its Hilbert space can be by a composition of qutrits, three basis states of which $\{0,A,B\}$ correspond to an unwritten register, the tick of $A$ and the tick of $B$ respectively,
\begin{align}
    \mathcal{H}_T = \mathcal{H}_{T_1} \otimes \mathcal{H}_{T_2} \otimes ... \ \text{with} \ \mathcal{H}_{T_n} = \mathbb{C}^3.
\end{align}
We assume that at $t=0$ no ticks have been recorded, and hence the register starts out with all qutrits initialized in the 0 state, $\bigotimes_n\ket{0}_{T_n}$. This modification does not affect the physics of the clocks, only the manner in which ticks are recorded.

The dynamics are modified accordingly. All operators previously containing shifts in individual registers of each clocks are replaced by operators where these shifts write the label of the ticked clock upon the first unwritten register: $\Gamma_{T_A} \otimes \mathds{1}_{T_B} \longrightarrow \Theta^{(A)}_{T}$,
\begin{align}
    \Theta^{(A)}_{T} &= \sum_{n=1}^\infty \left( \prod_{m=1}^{n-1} \left(\mathds{1} - \ket{0}\!\bra{0} \right)_{T_m} \right) \otimes \ket{A}\!\bra{0}_{T_n} \otimes \mathds{1}_{\text{rest}},
\end{align}
and analogously for $B$. This form of the operator allows to ignore the part of the register already written upon, write the label of the clock upon the first unwritten slot, and leave the rest unchanged.

\begin{figure}[t]
    \centering
    \includegraphics[width=.49\textwidth]{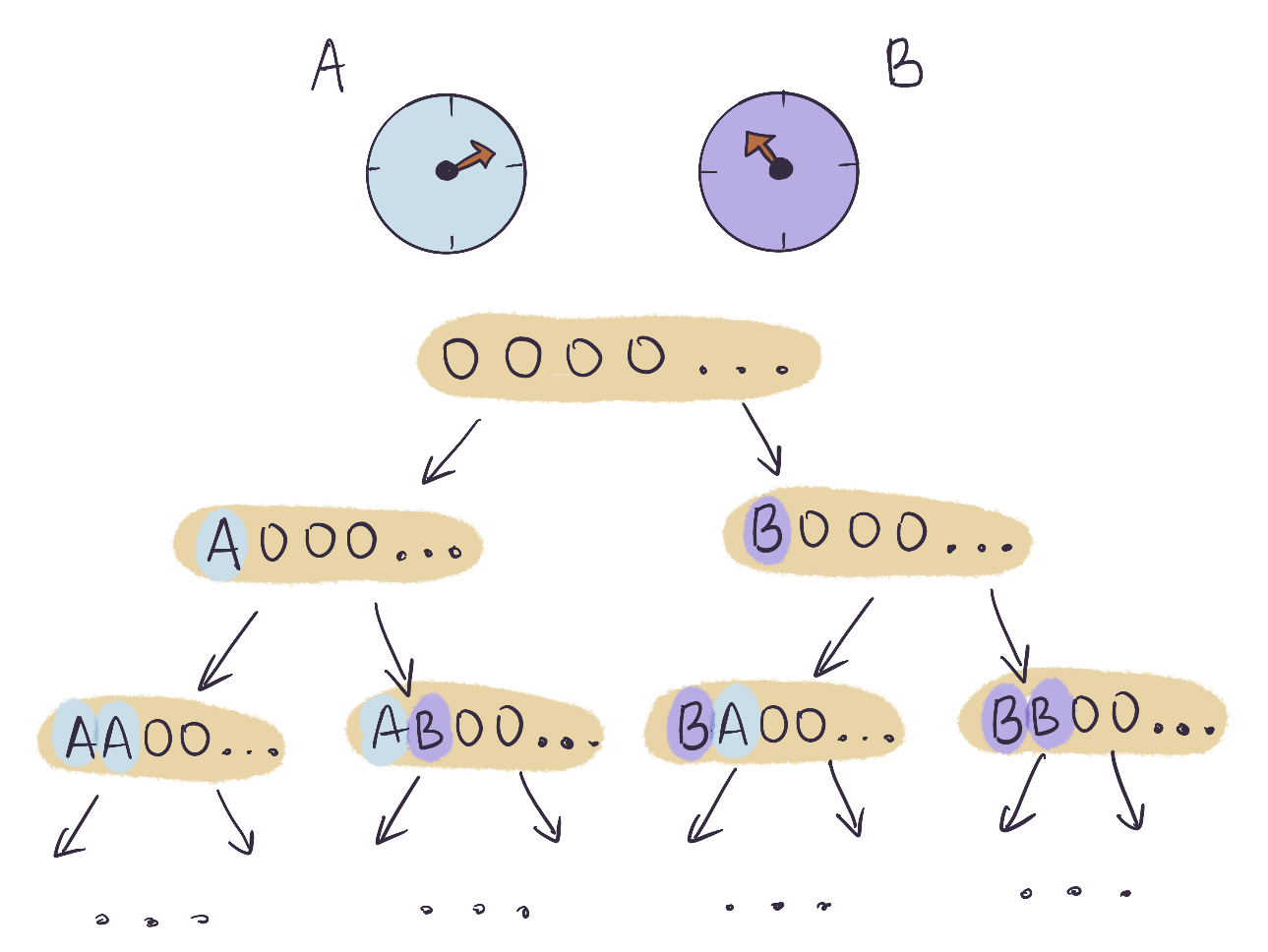}
    \caption{{\bf Multiple clocks writing a tick sequence upon a shared register.} All possible tick sequences observed for two clocks $A$ and $B$ can be represented as a tree, where each new level indicates the number of observed ticks increasing by one.}
    \label{fig:ticksequences}
\end{figure}

The result is illustrated in Fig.~\ref{fig:ticksequences}. 
The register state space is best visualised as a tree, with the initial node being the unwritten sequence, which then branches out, each branch corresponding to the next tick being $A$ or $B$. At any time $t>0$, the register reflects the order of all the previous ticks from $A$ and $B$.

By observing the tick sequence we can easily reconstruct the statistics $p(n_A|t),p(n_B|t)$ of the individual clocks by counting the total number of ticks of each clock within the sequence. However, the information contained in the tick sequence picture is strictly greater as it also includes the ordering between ticks.

An elegant feature of the KI decomposition for clocks is that it includes this structure of multiple clocks and a tick sequence. Namely, the structure of Hilbert space and dynamics above satisfies the KI structure, with the only difference from an elementary clock being that the states of the register that the clocks move between are not arranged in a serial chain, but rather as a tree, see Fig.~\ref{fig:ticksequences}.
Thus playing ticking clocks vs one another to form a tick sequence --- which is necessary to form an \textit{operational} theory of ticking clocks --- is seen to be not a separate model but rather another instance of a ticking clock, with a non-serial register.

\subsection{Abstract vs operational information revisited} 

A key observation in the section on the information content of a single clock was that the objects were \textit{abstract}: as they are all defined w.r.t. background time $t$ they are not directly observable. 
An operational information theory can only be based on comparing clocks to each other which is precisely the tick sequence we have just discussed. 
One can build operational analogues to abstract objects --- a simple example is $P(T_n = t)$, the probability density that the $n^{th}$ tick of the clock arrives at $t$. 
The operational analog of this would be to have a second clock $B$, and when the $n^{th}$ tick of $A$ occurs one observes the number of ticks so far from $B$ rather than background time, corresponding to a probability distribution $P(N_B^{(n)} = m)$. 
This does not need to be measured immediately on ticking, the advantage of the tick sequence is that the information once recorded can be observed at any later time.

Developing an information theory of tick sequences is worth a research work on its own and is the main focus of upcoming work~\cite{InfoClock}. One immediate question given the abstract vs operational distinction is whether one must lose information going from the former picture to the latter, and if so, how much. 
Clearly if we know the behaviour of a set of clocks $\{C_j\}_j$ w.r.t. background time $t$ --- from their times of arrival $P_j(T_1 = t_1,T_2 = t_2,...)$ for instance --- then we can recover all of the tick sequences that they would generate together. 
The converse is not obviously true and the gap between the two is an important question. 
A positive result in that direction is~\cite{RefClock}, where it is shown that, using arguably the simplest reference clock (a Poisson process), one can recover the abstract waiting-time distribution of an unknown reset clock from the tick sequence it generates with the reference.

\section{Previous literature}\label{sec:literature}

\subsection{Independence vs a classical register}


The bipartite structure of a ticking clock featuring a classical register is not a novel concept; for example, it is an integral part of the models presented in~\cite{rankovic_quantum_2015,erker_quantum_2014}, where ticking clocks are studied from the viewpoint of quantum information theory. There, the authors argue that for a clock that marks the passage of time in an \textit{operational} manner -- that is, the time read off the clock is interpreted as its measurable output -- the time information must be written upon a classical register. The bipartite structure of the clock is also implicit in the models of~\cite{woods_quantum_2022,erker_autonomous_2017} and is postulated in an explicit manner in~\cite{woods_autonomous_2021}, where the author also discusses an axiomatic framework for ticking clocks.


If one starts out by assuming the bipartite structure, then the independence of the clock --- i.e. the fact that the clock is not disturbed by time measurements --- follows immediately. This is the opposite direction to the path we have taken in this paper, where we postulate the independence principle as fundamental, and the structure is then implied by it.


At first glance this may appear to be a redundant chicken-and-egg conundrum: why should one favour either of the above as a starting point? Here we argue why independence is indeed a more appropriate fundamental starting point.

Firstly, there are a number of implicit assumptions in the model of an elementary clock that are exposed in detail here. The KI decomposition of the clock does not feature an explicit system that can be called a register, rather only an index corresponding to different states of a display. The explicit `register' system in an elementary clock is an an abstraction of this degree of freedom. This insight is neither deep nor novel, but it is missed when one begins already with an explicit register. In the same light, every real clock has a `display' system whose dynamical behaviour does not participate in the time-keeping properties of the clock. Again, this is entirely redundant while discussing the information theory of clocks, but it is elegant to have the true form of the register and display appear in a construction from first principles.

Moreover, by beginning from the independence principle we have that ticking clocks must be `emergent': i.e. that we can only have dynamics that maintains the classical nature of the register as a coarse-grained approximation. Under fully unitary dynamics, no such dynamics could exist. Thus every ticking clock has a limit to the frequency with which time-information can be gleaned from it, and cannot be continuously observed, contrary to the statement of~\cite{woods_autonomous_2021}. 

This frequency limit comes as no surprise, as similar coarse-graining assumptions, such as Born, Markov and rotating wave approximations, are required for the derivation of master equations describing dynamics of open quantum systems. If any of these assumptions is dropped, the reduced description of the system's dynamics will be rendered erroneous. It is thus important for an axiomatic approach to justify or derive these assumptions from first principles. 

Our approach also allows us to first arrive to a description of general ticking clocks that is broader than the usual model, the elementary clock. As we have pointed out, this is necessary in order to include some time-keeping systems that we would like to take into the consideration. In turn, each of the four additional properties needed to reduce our description to the elementary clock can be seen as a physical restriction that one can choose rather than a technical necessity for a ticking clock. As a result, we account for how the theory changes with each simplification.


For example, our construction does not apriori assume the irreversibility of ticks, which is a key assumption in all previous quantum-informational models of clocks~\cite{rankovic_quantum_2015,woods_quantum_2022,erker_autonomous_2017,yang_accuracy_2019,yang_ultimate_2020,woods_autonomous_2021,schwarzhans_autonomous_2021,meier_fundamental_2023}, as opposed to the thermodynamic models~\cite{barato_thermodynamic_2015,barato_cost_2016} that have clocks with reversible ticks. This distinction is linked to the entropy production associated with macroscopic irreversiblity. We note that in both cases, the asymptotic behaviour of the clock is accounted for by full counting statistics, weakening the distinction between them.



\subsection{Bounds on the performance of clocks}

A clock is, ultimately, a tool that can be used in any operation requiring to keep track of time, and one of its defining properties is its precision, which characterizes its performance. Here we list some of the existing results and open questions regarding the bounds on precision of clocks:

\begin{itemize}
    \item \textbf{Fundamental thermodynamic / entropic cost.} Refs.~\cite{barato_cost_2016,erker_autonomous_2017,milburn_thermodynamics_2020} suggest a trade-off between a clocks precision and its entropy dissipation, albeit for a restricted class of clocks. Proving a general trade-off would represent understanding the arrow of time from the perspective of the devices that actually measure said arrow. Promising directions in this regard include the promising fields of thermodynamic and kinetic uncertainty relations, as also information-theoretic approaches such as~\cite{yang_ultimate_2020};
    \item \textbf{Dimension of clockwork space.} \cite{woods_quantum_2022} proves a bound on the precision of clockworks with incoherent dynamics as a function of the Hilbert space dimension of the clockwork, and that there exist quantum clockworks that outperform them. Subsequently,~\cite{yang_ultimate_2020} proved a bound for quantum clockworks as well; the quantum clock has a quadratic advantage over a classical one w.r.t. the dimension of the clockwork Hilbert space. While the optimal construction of the classical (incoherent) clockworks is known, the optimal quantum clockworks remain an interesting open question;
    \item \textbf{Precision vs. frequency trade-off.} In~\cite{meier_fundamental_2023} the authors take the tick dynamics of the clock to be fundamental (represented by the rates $\gamma_j$ attached to the Lindblad jump operators that move the register), and demonstrate that in this case the precision of the clock can only be increased by sacrificing its frequency of ticking. In~\cite{schwarzhans_autonomous_2021}, the trade-offs are shown for a clock the transitions of which are driven by a thermal engine. The quantum clock construction in~\cite{woods_quantum_2022} satisfies the scaling of said trade-off, and it remains open as to whether there exists a quantum over classical advantage here too.
\end{itemize}

\section{Discussion \& Outlook}\label{sec:outlook}

\subsection{General vs elementary clocks}

In this paper, we derive a structure for ticking clocks that is more general than the usual model in the literature, namely elementary clocks. It is not clear whether the additional restrictions in the elementary model have a significant influence on the achievable performance of the clocks. For completeness, here we recap all open questions related to this loss of generality mentioned in the main text.

\begin{itemize}
\item  \textbf{Self-timing.} Even the best clocks --- atomic clocks --- are not strictly self-timed, rather their dynamics exhibit a frequency drift.  Modelling the long term behaviour of such systems, i.e. with drifting dynamics is necessary to connect the research on ticking clocks to practical implementations.

\item \textbf{Clockwork independence.} It remains an open question whether one can obtain greater precision by using different dynamics for different ticks. The results of \cite{woods_quantum_2022} suggest that this is impossible in the case that the clockwork dynamics is incoherent; the argument is that the precision over many ticks of such a clock is bound by the sum of the precision of each tick in turn. It is unclear whether quantum clockworks allow one to circumvent this bound.

\item \textbf{Serial registers.} The same arguments from \cite{woods_quantum_2022} suggest that given a network of register states, the precision of moving from one state to another is maximised when there is only one possible path rather than a convex mixture of the same. This picks out the optimal register as a chain of states with only nearest neighbour jumps. Proving this in generality would be useful; if it is indeed not the case, then there would exist more precise clock constructions that employ the ability to jump many steps on the register.

\item \textbf{Reversible ticks.} There are indications that reversible ticks cannot increase the precision of clocks. An example is Ref.~\cite{barato_cost_2016} where they posit a bound on the precision of such clocks that is highest when the clock only ticks in one direction. Proving this also remains an interesting open question.
\end{itemize}

\subsection{Thermodynamic cost of clocks}

Our starting point for this paper was the fact that clocks --- specifically those apparently undisturbed by measurement --- exist, and our goal was to model them.

A property that holds for all ticking clocks is that they are out-of-equilibrium systems, and are maintained so. This implies that thermal resources must be expended in the working of any clock, as noted in the previous section.

In quantifying this connection one needs to precisely define the notion of an \textit{autonomous clock}, i.e. a model where the resources involved in keeping the clock out of equilibrium appear explicitly and can be accounted for. While it is typically straightforward to argue whether a specific model is or is not autonomous, a general criterion does not exist in the case of ticking clocks with a quantum clockwork\footnote{For a clockwork with only incoherent dynamics, it is arguably autonomous \textit{iff} the Lindbladian generator satisfies local detailed balance.}. An important task is thus to derive the general construction of an autonomous quantum ticking clock, both in and beyond the weak-coupling limit.

\subsection{The emergent continuity of classical clocks}

Consider a classical system such as a pendulum. This can be interpreted as a clock: by measuring the angle of the pendulum from the vertical one obtains information about time (modulo the time period).

This system appears to contradict a central conclusion of our work: on the one hand it appears to be an independent clock undisturbed by us looking at it, on the other hand the time information --- the angle $\theta$ --- is continuous rather than than taken from a discrete set; thus not a ticking clock.

This apparent continuity is however only an approximation. Observing the angle of the pendulum involves interacting with photons of a particular wavelength. The wavelength sets the imprecision $\delta \theta $ of the measurement. This implies that the time information is effectively discrete: one can divide the range of $\theta$ into intervals of $\delta \theta$. For true continuity one would require vanishing wavelength, which in turn implies photons of unbounded energy that would disturb the motion of the pendulum.

In a similar manner, every classical clock is thus also a discrete ticking clock, with the property that the measurement precision is high enough to make the time information appear continuous while low enough as to not disturb the clock.

\section{Acknowledgements.}

The authors acknowledge insightful discussions with Paul Erker, Marcus Huber, Maximilian Lock, Florian Meier, Renato Renner, Emanuel Schwarzhans and Mischa Woods. We also thank Florian Meier for pointing out the connection to full counting statistics. R.S. acknowledges funding from the Swiss National Science Foundation via an Ambizione grant PZ00P2\_185986. N.N. acknowledges support from the Swiss National Science Foundation through SNSF project No. 200021\_188541 and through the the National Centre of Competence in Research Quantum Science and Technology (QSIT). H.W. acknowledges support by the DFG through SFB 1227 (DQ-mat), Quantum Valley Lower Saxony, and funding by the Deutsche Forschungsgemeinschaft (DFG, German Research Foundation) under Germany's Excellence Strategy EXC-2123 QuantumFrontiers 390837967. 

\bibliographystyle{apsrev4-2}
\bibliography{bib-clocks}

\clearpage
\appendix

\section*{Appendix}
\section{Koashi-Imoto decomposition}
\label{appendix:KI}

Below we discuss the theorem postulating Koashi-Imoto decomposition. The description (adapted to our settings) is borrowed from~\cite{hayden_structure_2004}, the results of which are identical to those of Koashi-Imoto~\cite{koashi_operations_2002}. Refs.~\cite{jencova_sufficiency_2006,kuramochi_accessible_2018} show that the results also transfer to infinite-dimensional Hilbert-spaces.

Suppose that we have a set of density matrices $\{\rho_k\}_k \in L(\mathcal{H})$, and want to find the structure of the set of quantum channels (trace-preserving and completely positive linear maps) $\mathcal{E}:L(\mathcal{H})\to L(\mathcal{H})$ which leave these states invariant,
\begin{align}
    \mathcal{E}(\rho_k) = \rho_k \ \forall k.
\end{align}
According to the Stinespring dilation theorem, each such channel admits a representation of a partial trace over a unitary evolution of the system $S$ together with the environment,
\begin{align}
    \mathcal{E}(\rho_S) = \Tr \left(U_{SE} (\rho_S\otimes\sigma_E)U_{SE}^\dagger\right),
\end{align}
as well as a decomposition in terms of Kraus operators,
\begin{align}
\label{eq:kraus}
    \mathcal{E}(\rho_S) = \sum_j A_j\rho_S A_j^\dagger,
\end{align}
with $\sum_j A_j^\dagger A_j = \mathds{1}$.
The Koashi-Imoto decomposition then implies that the Hilbert space of the system can be written as
\begin{align}
    \mathcal H = \bigoplus_n \mathcal{H}_{C_n} \otimes \mathcal{H}_{F_n},
\end{align}
and the states w.r.t. this decomposition look like
\begin{align}
    \rho_k &= \bigoplus_n p_{n|k} \; \rho^{(n)}_k \otimes \omega_n.
\end{align}
Moreover, for every quantum channel $\mathcal{E}$ which leaves the states $\rho_k$ invariant, the associated unitary has the form
\begin{align}
    U_{SE} = \bigoplus_n \mathds{1}_{C_n} \otimes U_{F_n E}
\end{align}
with unitaries $U_{F_n E}$ acting on $\mathcal{H}_{F_n}\otimes \mathcal{H}_E$ that don't change the state $\omega_n$,
\begin{align}
    \Tr\left(U_{F_n E} (\omega_n\otimes\sigma_E)U_{F_n E}^\dagger\right)=\omega_n.
\end{align}
The above defines a quantum channel on states of subspaces $\mathcal{H}_{F_n}$, and can be described via Kraus decomposition, 
\begin{align}
    \sum_j A_{j,n} \omega_n A_{j,n}^\dagger = \omega_n.
\end{align}
This allows us to write down the explicit form of Kraus operators $A_j$  in~\ref{eq:kraus} as
\begin{align}
    A_j &= \bigoplus_n \mathds{1}_{\mathcal{C}_n} \otimes A_{j,n}.
\end{align}

\section{Elementary reset clocks: connecting asymptotic rates to moments of the waiting time}\label{app:cumulantsvswaitingtime}

The probability of the number of ticks at time $t$ being equal to $n$ is
\begin{align}
    p_{n|t} = F_{n|t} - F_{n+1|t},
\end{align}
where $F_{n|t}$ is the cumulative distribution function (CDF) of the waiting time of the $n^{\mathrm{th}}$ tick. For elementary reset clocks this is the $n$-fold convolution of the waiting time distribution $\omega_t$ for a single tick, 
\begin{align}
    F_{n|t} = \int_0^t \underbrace{\omega*\dots*\omega_{t^\prime}}_{n} dt'.
\end{align}
From this expression, we can derive the expectation value of $n^k$ ($k$th moment in the number of ticks), $k \geq 1$,
\begin{align}
    \braket{n^k}_t &= \sum_{n=0}^\infty n^k p_{n|t} \\
    &= \sum_{n=0}^\infty n^k \left(F_{n|t} - F_{n+1|t} \right) \\
    &= \sum_{n=1}^\infty \left(n^k - (n-1)^k\right) F_{n|t} \\
    &= \sum_{n=1}^\infty \sum_{l=0}^{k-1} \binom{k}{l} (-1)^{k-l+1} n^l F_{n|t}.
\end{align}
Now we apply the Laplace transform to both sides of the equation, which maps all convolutions to the powers of Laplace transforms of single ticks waiting-time distributions (we denote the Laplace transform of a function by adding a superscript $\cdot^*$ and the change of variable $t \mapsto s$),
\begin{align}
    F_{n|t} \mapsto \frac{\left(\omega^*_s\right)^n}{s}.
\end{align}
Thus, we arrive to 
\begin{align}
    \langle N^k \rangle^*_s = \sum_{n=1}^\infty \sum_{l=0}^{k-1} \binom{k}{l} (-1)^{k-l+1} n^l \frac{\left(\omega^*_s\right)^n}{s}.
\end{align}
Now we use the formula for calculating the sum $\sum_{n=1}^\infty n^k x^n$,
\begin{align}
    \sum_{n=1}^\infty n^k x^n = \frac{x\cdot A_k(x)}{(1-x)^{k+1}},
\end{align}
where $A_k(x)$ are Eulerian polynomials defined recursively,
\begin{align}
    A_0(x) &=0; \\
    A_k(x) &= x(1-x) A_{k-1}'(x) + A_{k-1}(x) \left(1 + (k-1)x\right).
\end{align}
This allows us to rewrite 
\begin{align}
    \langle N^k \rangle^*_s &= \frac{1}{s} \sum_{l=0}^{k-1} \binom{k}{l} (-1)^{k-l+1} \sum_{n=1}^\infty  n^l \left(\omega^*_s\right)^n \\
    &= \frac{1}{s} \sum_{l=0}^{k-1} \binom{k}{l} (-1)^{k-l+1}  \frac{\omega^*_s \cdot A_l\left(\omega^*_s\right)}{(1-\omega^*_s)^{l+1}}.
\end{align}
To find the asymptotic behaviour, we expand the Laplace transform of the single tick distribution in terms of its moments,
\begin{align}
    \omega^*_s = 1 - \mu_1 s + \frac{\mu_2 s^2}{2} - \dots,
\end{align}
and apply the inverse Laplace transform, which maps
\begin{align}
    \frac{1}{s^j} \mapsto \frac{t^{j-1}}{(j-1)!}, \ j\in\mathbb{N}.
\end{align}

\textbf{Asymptotics of the first two moments.} 
For the expected number of ticks at time $t$ and its Laplace transform approximation, we obtain
\begin{align}
    \langle n \rangle_t &= \sum_{n=1}^\infty F_{n|t}; \\
    \langle n \rangle^*_s &= \frac{1}{s} \sum_{n=1}^\infty \left(\omega^*_s\right)^n \\ 
    &= \frac{\omega^*_s}{s(1-\omega^*_s)} \\
    &\approx \frac{1}{\mu_1 s^2}\left(1-\mu_1 s + \frac{\mu_2 s}{2\mu_1} + O(s^2)\right).
\end{align}
Applying the inverse transform, we get the desired asymptotic rate (current of ticks)
\begin{align}
    \braket{n}_t &\to \frac{t}{\mu_1} + \left(\frac{\mu_2}{2\mu_1^2} - 1\right), \\
    \lim_{t \to \infty} \frac{\braket{n}_t}{t} &= \frac{1}{\mu_1}.
\end{align}

We perform a similar analysis for the second moment, ending with the asymptotic behaviour:
\begin{align}
    \langle n^2 \rangle_t &\to \frac{t^2}{\mu_1^2} + \left(\frac{2(\mu_2-\mu_1^2)}{\mu_1^3} - \frac{1}{\mu_1}\right) t \nonumber \\
    &\quad\quad + \left(\frac{3\mu_2^2}{2\mu_1^4} - \frac{2\mu_3}{3\mu_1^3} - \frac{\mu_2}{\mu_1^2}\right),
\end{align}
so that the variance (second cumulant) is linear in $t$ for large times,
\begin{align}
    \langle n^2 \rangle_t - \langle n \rangle^2_t &\to \left(\frac{\mu_2}{\mu_1^3} - \frac{1}{\mu_1}\right)t + \text{const} \\
    &= \frac{\sigma^2}{\mu_1^3} t + \text{const}.
\end{align}
where $\sigma^2$ is the variance of the waiting-time $\omega_t$.

The ratio of asymptotic rates thus satisfies the condition stated in the main text:
\begin{align}
    R_1 &= \lim_{t \to \infty} \frac{d_t \braket{n}_t}{d_t \left( \braket{n^2} - \braket{n}^2 \right)} = \frac{\mu^2}{\sigma^2} = R_2.
\end{align}

\section{Asymptotic Allan variance for self-timed clocks}
\label{appendix:allan-variance}

Allan variance is used as a measure of frequency stability in clocks and similar devices, which require signal regularity. To calculate it in the simplest setting without a dead-time between measurements, one partitions the total observation time $T$ into $M$ bins of length $\tau$ (the averaging time). One then then determines for each bin ("sample") the average frequency and takes the average over the collection of $2$-sample variances.
For simplicity, let us assume that we start the sampling process for frequencies $\nu$ at $t=0$. The first 2-sample variance is defined by
\begin{align}
    \sigma^2_{\nu,1} &= \left(\nu_1 - \frac{\nu_1+\nu_2}{2}\right)^2 +\left(\nu_1 - \frac{\nu_1+\nu_2}{2}\right)^2 \\ 
    &= \frac{(\nu_1-\nu_2)^2}{2},
\end{align}
where $\nu_1$ and $\nu_2$ are the observed average frequencies in the time intervals $[0,\tau]$ and $[\tau, 2\tau]$ respectively. In terms of the numbers of observed ticks at times $0,\tau$ and $2\tau$, they can be written as 
\begin{align}
    \nu_1 &= \frac{n_\tau - n_0}{\tau}, \\
    \nu_2 &= \frac{n_{2\tau} - n_\tau}{\tau}. 
\end{align}
Substituting into the 2-sample variance above, we obtain
\begin{align}
    \sigma^2_{\nu,1} &=\frac{1}{2\tau^2}\left(n_{2\tau} - 2n_\tau + n_0\right)^2.
\end{align}

We repeat this procedure for each consecutive pairs of intervals $\left[k\tau,(k+1)\tau\right]$ and $\left[(k+1)\tau,(k+2)\tau\right]$ for $k\in\{0,\dots,M-1\}$. The average over the expectation value of all 2-sample variances gives us the Allan variance $A$ of the process,
\begin{align}
    A &= \frac{1}{M}\sum_{k=0}^{M-1} \langle \sigma^2_{\nu,k}\rangle \\
    &= \frac{1}{M}\sum_{k=0}^{M-1} \frac{1}{2\tau^2}\langle\left( n_{(k+2)\tau} - 2n_{(k+1)\tau} + n_{k\tau}\right)^2\rangle. \label{eq:av-main}
\end{align}

We can calculate the above if we assume that the clock is in its unique steady state, i.e. in the asymptotic limit $t\rightarrow \infty$.
In this limit, the two-time correlations appearing in~\ref{eq:av-main} can be calculated using the quantum regression theorem~\cite{schaller_technical_2014},
\begin{align}
    \frac{d}{d\tau} \langle n_t n_{t+\tau}\rangle &= \frac{1}{\mu}\langle n_t \id_{t+\tau}\rangle = \frac{t}{\mu^2} \quad (t\gg 1).
\end{align}
The boundary condition for $\tau=0$ reads $\langle n_t^2\rangle = \frac{t^2}{\mu^2} + \Sigma t$, which follows from the general asymptotic properties of Markovian processes (see Sec.~\ref{subsec:counting-statistics}) and results in the solution
\begin{align}
    \langle n_t n_{t+\tau}\rangle &= \langle n_t^2 \rangle + \frac{t\tau}{\mu^2} = \frac{t(t+\tau)}{\mu^2} + \Sigma t.
\end{align}
Calculating the Allan variance, we obtain
\begin{align}
    A &= 
    \frac{1}{M} \sum_{k=0}^{M-1} \frac{\Sigma}{\tau} = \frac{\Sigma}{\tau}.
\end{align}

\section{Continuity of the set of measurement outcomes for a non-independent clock}\label{appendix:discretetick}

A simple clock construction due to Salecker and Wigner~\cite{salecker_quantum_1958}, later expounded by Peres~\cite{peres_measurement_1980}, is the following Hamiltonian: an equally spaced, non-degenerate and finite spectrum,
\begin{align}
    H &= \sum_{n=0}^{d-1} n \omega \ket{n}\!\bra{n}.
\end{align}

Its likeness to a clock is seen through the evolution of the following basis of states:
\begin{align}
    \ket{\theta_k} &= \frac{1}{\sqrt{d}} \sum_{n=0}^{d-1} e^{-i 2\pi n k /d} \ket{n},
\end{align}
where $k \in \{0,1,2,...,d-1\}$ (in the following $k \equiv k \mod d$). One finds that every `angle' state $\ket{\theta_k}$ evolves to $\ket{\theta_{k+1}}$ in the time period $2 \pi/(\omega d)$,
\begin{align}
    e^{- i H \frac{2 \pi}{\omega d} }\ket{\theta_k} &= \ket{\theta_{k+1}},
\end{align}
Hence the label `angle states', they may be arranged on a circular clock face, each $\ket{\theta_k}$ at an angle of $2\pi k/d$.

A simple manner of employing this system as a clock would be to initialise it in the state $\ket{\theta_0}$ at $t=0$, and measure in the $\theta$ basis at a later time, corresponding the set of projectors and corresponding outcomes:
\begin{align}
    \Pi_k &= \ket{\theta_k}\!\bra{\theta_k}, \quad t_k = \frac{2 \pi k}{\omega d}.
\end{align}

At this point it appears that the clock has a discrete structure in that it measures time in terms of a fundamental unit. However we are free to pick another set of angle states: take $\alpha \in  (0,1)$ and define a new set via
\begin{align}
    \ket{\lambda_k} &= \frac{1}{\sqrt{d}} \sum_{n=0}^{d-1} e^{-i 2\pi n (k+\alpha) /d} \ket{n}
\end{align}
For every $\alpha$ the above is an orthonormal basis, and one can check that $\ket{\theta_k}$ evolves into $\ket{\lambda_k}$ in a time step of $\frac{2\pi \alpha}{\omega d}$. Thus the new set is shifted w.r.t. the original one by the angle $\frac{2\pi \alpha}{\omega d}$.

We can now construct a measurement that incorporates the new basis as well, for instance the following set of $2d$ POVM elements:
\begin{align}
    \Pi_k &= \frac{1}{2} \ket{\theta_k}\!\bra{\theta_k}, \quad t_k = \frac{2 \pi k}{\omega d}, \\
    \Pi^\prime_l &= \frac{1}{2} \ket{\lambda_l}\!\bra{\lambda_l}, \quad t_l = \frac{2 \pi (l+\alpha)}{\omega d}.
\end{align}
Indeed one can generalise to an arbitrary number of values of $\alpha \in [0,1)$. Thus the time measured by this clock does not have a fundamental unit, even though each individual measurement might have a finite number of outcomes.

\end{document}